\documentclass[twocolumn,aps,prl,showpacs,amsmath,amssymb,superscriptaddress]{revtex4}

\usepackage[usenames]{color}

\usepackage{graphicx}

\def\<{\langle}
\def\>{\rangle}

\begin{document}

\title{Dynamics of three-body correlations in quenched unitary Bose gases}

\author{V. E. Colussi}
\affiliation{Eindhoven University of Technology, PO Box 513, 5600 MB Eindhoven, The Netherlands}
\affiliation{JILA, NIST and Department of Physics, University of Colorado, Boulder, Colorado 80309-0440, USA}
\author{J. P. Corson}
\affiliation{JILA, NIST and Department of Physics, University of Colorado, Boulder, Colorado 80309-0440, USA}
\author{J. P. D'Incao}
\affiliation{JILA, NIST and Department of Physics, University of Colorado, Boulder, Colorado 80309-0440, USA}

\begin{abstract}
We investigate dynamical three-body correlations in the Bose gas during the earliest stages of evolution after a quench to the unitary regime.  The development of few-body correlations is theoretically observed by determining the two- and three-body contacts.  We find that the growth of three-body correlations is gradual compared to two-body correlations.  The three-body contact oscillates coherently, and we identify this as a signature of Efimov trimers.  We show that the growth of three-body correlations depends non-trivially on parameters derived from both the density and Efimov physics.  These results demonstrate the violation of scaling invariance of unitary bosonic systems via the appearance of log-periodic modulation of three-body correlations.  
\end{abstract}


\maketitle 
 {\it Introduction.}  In the ultracold regime of bosonic gases, where the interaction is well-characterized by the $s$-wave scattering length, $a$, macroscopic theories of matter can be formulated from microscopic Hamiltonians.  These theories relate the physics of a few atoms to their manifestations in the macroscopic observables.  At the heart of this link is a set of universal relations due to Shina Tan \cite{TAN20082952,TAN20082971,TAN20082987}.  The Tan relations provide an alternative path to calculate thermodynamical properties of an ultracold quantum gas by simply studying analytic solutions of the two-body problem and extracting the extensive two-body contact density $\mathcal{C}_2$, which characterizes two-body correlations at short distances within the system.  These relations are well-understood for two-component Fermi gases even in the unitary regime $n|a|^3\gg1$, where $n$ is the atomic density, and have been verified experimentally \cite{PhysRevLett.105.070402,PhysRevLett.104.235301}.  For strongly-interacting Bose gases there is an additional complication due to the existence of the three-body Efimov effect where an infinity of three-body trimers emerges at unitarity $(a=\infty)$ that strongly alters scattering observables at ultracold energies \cite{efimov1979low,BRAATEN2006259,WANG20131,d2017few}.  Here, universal relations between few-body physics and macroscopic observables involve also the three-body contact density $\mathcal{C}_3$ \cite{PhysRevLett.106.153005,PhysRevA.86.053633}, central to the many-body theory, and whose properties are not yet theoretically known for the degenerate Bose gas in the unitary regime.      

Unlike two-component unitary Fermi gases, strongly-interacting Bose-condensed gases are plagued by an enhanced three-body loss rate growing as $n^3 a^4$, limiting the development of correlations.  By quenching the interactions from weak to unitarity, Makotyn {\it et al.} \cite{makotyn2014universal} observed saturation of the single-particle momentum distribution---an observable sensitive to few-body correlations---of the quenched unitary degenerate Bose gas on a timescale shorter than the observed atom loss rate.  It has been suggested that the observed tail of the saturated momentum distribution oscillates log-periodically, the signature of Efimov physics, and therefore is a measurement of nonzero $\mathcal{C}_3$ \cite{PhysRevA.92.062716,PhysRevLett.112.110402}.  For the thermal unitary Bose gas, $\mathcal{C}_3$ has been directly measured interferometrically in Fletcher {\it et al.} \cite{Fletcher377} approaching the theoretical saturation value from Ref.~\cite{PhysRevLett.112.110402}.

The introduction of additional length scales due to Efimov physics can break the continuous scale invariance of system properties with the interparticle spacing $n^{-1/3}$.  That all properties of unitary quantum gases depend solely on the interparticle spacing is referred to as the universality hypothesis \cite{PhysRevLett.92.090402}.  Within this hypothesis, for bosons or fermions, the only relevant scales are set by the momentum $\hbar k_n=\hbar (6\pi^2 n)^{1/3}$, energy $E_n=\hbar^2k_n^2/2m$, and time $t_n=\hbar/E_n$ where $m$ is the atomic mass.  Although $\mathcal{C}_2$ within the non-equilibrium regime is well-studied \cite{PhysRevA.89.021601,PhysRevA.91.013616,PhysRevA.88.063611}, predicting the corresponding time evolution, scaling properties, and saturation value of $\mathcal{C}_3$ is still an open problem, limiting our full understanding of the role of Efimov physics in the quenched unitary Bose gas.

In this Letter, we theoretically observe the growth of the {\it dynamical} three-body contact density $\mathcal{C}_3$ immediately following the quench to unitarity.  We have developed a simple model that describes the early correlation dynamics of the quenched unitary Bose gas using known analytic solutions of the three-body problem \cite{PhysRevLett.97.150401}.  At the earliest stages of evolution, we find that the three-body contact grows slowly compared to the two-body contact and exhibits coherent oscillations at the frequency of Efimov trimers.  Our results demonstrate the violation of the continuous scale invariance of $\mathcal{C}_3$ at early-times maximized whenever the size of an Efimov trimer is comparable to the interparticle spacing.  

{\it Relations at short distances.}  We begin by establishing the short distance connections between the two- and three-body correlations of a Bose gas, the two- and three-body contacts, and solutions of the few-body problem.  These connections are made at distances larger than the van der Waals length, $r_\mathrm{vdW}$, but smaller than other length scales of the problem ($a$, $n^{-1/3}$, etc.)  Here this is done for a uniform gas of $N$ particles in volume $V$ with density $n=N/V$, which can be generalized simply to trapped gases by using the local-density approximation, in which case $n$ is the average density $\langle n \rangle$.

Within the zero-range model for the interatomic interactions, the short distance behavior of the two- and three-body correlation functions is determined exclusively by the two- and three-body contacts (see Ref.~\cite{PhysRevA.86.053633} and references therein)   
\begin{eqnarray}
&&g^{(2)}({\bf r,t})\underset{{|\bf r|}\rightarrow 0}{=}\frac{1}{16\pi^2 n^2 r^2}\mathcal{C}_2,\label{eq:g2c2}\\
&&g^{(3)}(R,{\bf \Omega},t)\underset{{R}\rightarrow 0}{=}|\Psi_\mathrm{sc}(R,{\bf \Omega})|^2\frac{8}{n^3 s_0^2\sqrt{3}} \mathcal{C}_3,\label{eq:g3c3}
\end{eqnarray}
where $s_0\approx \ 1.00624$ is Efimov's universal constant for three identical bosons \cite{efimov1971weakly}.  The center of mass dependence in the equations above has been suppressed due to translational invariance. The relative atomic configuration is parametrized by the Jacobi vectors ${\bf r}\equiv{\bf r_2-r_1}$ and ${\boldsymbol \rho}\equiv(2{\bf r_3}-{\bf r_1}-{\bf r_2})/\sqrt{3}$.  Alternatively, it can be parametrized by the hyperradius $R^2\equiv(r^2+\rho^2)/2$ and the set of hyperangles  ${\bf \Omega}=\{\alpha,\bf{\hat{r}},\boldsymbol{ \hat{ \rho}}\}$, containing the hyperangle $\alpha=\arctan (r/\rho)$ and spherical angles for each Jacobi vector.  The limit notation in Eqs.~(\ref{eq:g2c2})--(\ref{eq:g3c3}) indicates $|{\bf r}|\to0$ for fixed $\bf{\hat{r}}$, and $R\to 0$ for fixed ${\bf \Omega}$, respectively.  $\Psi_\mathrm{sc}(R,{\bf \Omega})$ is the zero-energy three-body scattering wave function
\begin{equation}
\Psi_\mathrm{sc}(R,{\bf \Omega})=\frac{1}{R^2}\sin \left[s_0\ln\frac{R}{R_t}\right]\frac{\phi_{s_0}({\bf \Omega})}{\sqrt{\langle\phi_{s_0}|\phi_{s_0}\rangle}},
\end{equation}
where $R_t/r_\mathrm{vdW}\in [1,e^{\pi/s_0}]$ is the three-body parameter, setting the phase of log-periodic oscillations, and $\phi_{s_0}({\bf \Omega})$ is the hyperangular wave function for three identical bosons in the state of lowest total angular momentum.  [For analytic expressions of $\phi_{s_0}({\bf \Omega})$ and the normalization constant $\langle\phi_{s_0}|\phi_{s_0}\rangle$, we refer the reader to Ref.~\cite{SM,PhysRevA.83.063614,PhysRevC.47.1876}.]   

After the interaction quench---amounting within our model to a quench of the Bethe-Peierls contact boundary condition at $r=0$---the contact dynamics occur exclusively at short distances.  Therefore, the short-time short-range behavior of few-body wave functions can yield quantitatively correct predictions for the contact dynamics of a quenched many-body system \cite{PhysRevA.91.013616,PhysRevA.94.023604}.  Generally, if a particle is measured at a location defining the origin of a coordinate system, then the quantity $n g^{(2)}({\bf r},t)$ is the  probability density for measuring another particle at ${\bf r}$ \cite{pathria}.  In a three-body model, that probability density is given in terms of the three-body wave function $\Psi({\bf r},{\boldsymbol\rho,t})$.  We are interested in this probability density at short distances where $\mathcal{C}_2$ is defined, suggesting the relation
\begin{equation}\label{eq:psig2link}
ng^{(2)}({\bf r},t)\underset{{|\bf r|}\rightarrow 0}{=}2\int d^3 r_{3,12}|\Psi({\bf r},{\boldsymbol\rho,t})|^2,
\end{equation}
where ${\bf r_{3,12}}={\boldsymbol\rho}\sqrt{3}/2$.  Additionally, the quantity $n^2 g^{(3)}(R,{\bf \Omega},t)$ is the probability density of finding two other particles at locations defined by the three-body configuration $(R,{\bf \Omega})$.  The analogous relation between the three-body correlation function and the three-body wave function is
\begin{equation}\label{eq:psig3link}
n^2g^{(3)}(R,{\bf \Omega},t)\underset{{R}\rightarrow 0}{=} 2 \ |\Psi(R,{\bf \Omega},t)|^2.
\end{equation}
The factor of $2$ in Eqs.~(\ref{eq:psig2link})--(\ref{eq:psig3link}) is due to the indistinguishability of the two particles not fixed at the origin.  

{\it Initial conditions.}  To make the links in Eqs.~(\ref{eq:psig2link})--(\ref{eq:psig3link}) quantitatively correct, we employ an unambiguous calibration scheme.  The three-body model yields correct short-time predictions of the contacts if and only if the initial wave function satisfies Eqs.~(\ref{eq:psig2link}) and (\ref{eq:psig3link}) at $t=0$.  For our model of the quenched gas, we start from the non-interacting limit where $g^{(2)}({\bf r},0)=g^{(3)}(R,{\bf \Omega},0)=1$.  There is considerable freedom in the choice of the initial three-body wave function satisfying these initial conditions.  Here, we choose 
\begin{equation}\label{eq:icpsi}
\Psi_0(R,{\bf \Omega})=Ae^{-R^2/2B_1^2}\left[1-\left(\frac{R}{B_2}\right)^2\right],
\end{equation}
where the analytic expression for the normalization constant $A$ is given in Ref.~\cite{SM}.  We find that setting $B_1\approx0.6009n^{-1/3}$ and $B_2\approx 1.1278 n^{-1/3}$ satisfies both initial conditions simultaneously.  With this calibration scheme, predictions for short-time short-distance correlation phenomenon for the quenched many-body system should not depend on the long-range part of the few-body wave function.  This point was demonstrated for one- and three-dimensions in Refs.~\cite{PhysRevA.91.013616,PhysRevA.94.023604}.    

{\it Three-body model at unitarity.}  After the quench to unitarity, the initial wave function [Eq.~(\ref{eq:icpsi})] is projected onto eigenstates at unitarity, for which we utilize solutions for three harmonically confined bosons first given by Werner and Castin in Ref.~\cite{PhysRevLett.97.150401}.  These eigenstates serve only as a convenient basis on which to expand the problem.  At unitarity within the zero-range model, the relative three-body eigenstates can be factorized as $\Psi_{s,j}(R,{\bf \Omega})=\mathcal{N}F^{(s)}_j(R)\phi_s({\Omega})/R^2\sin2\alpha,$ where $\mathcal{N}$ is a normalization factor, and $s$ is a solution of a transcendental equation resulting from the Bethe-Peierls contact condition taken at unitarity (see Ref.~\cite{SM}).  The hyperradial wave functions $F_j^{(s)}(R)$ obey \cite{PhysRevLett.97.150401} 
\begin{equation}\label{eq:hypeq}
\left[-\frac{\hbar^2}{2m}\left(\frac{d^2}{dR^2}+\frac{1}{R}\frac{d}{dR}\right)+U_s(R)\right]F^{(s)}_j(R)=EF_j^{(s)}(R),
\end{equation}
where $U_s(R)=\hbar^2s^2/(2mR^2)+m\omega_0^2R^2/2$ is a sum of the effective three-body potential in channel $s$ and of the local harmonic trap with frequency $\omega_0$ and trap length $a_\mathrm{ho}=\sqrt{\hbar/m\omega_0}$.  The index $j$ denotes a particular eigenstate within the channel, and $E$ is the three-body relative energy.  

To make a connection with the short distance behavior of three-body correlations, we need only consider the $R\to 0$ behavior of the hyperradial eigenstates. For $s>0$, the limiting behavior of the hyperradial eigenstates is $F_j^{(s)}(R)\propto \mathcal{O}(R^s)$, which does not contribute to the short-range three-body correlations.  The only channel that contributes to three-body correlations at short distances is the one associated with the imaginary solution of the transcendental equation denoted $s=is_0$, which gives rise to the attractive $1/R^2$ three-body potential that produces the Efimov effect.  The limiting behavior of the hyperradial eigenstates in the Efimov channel is $F_j^{(s_0)}\propto\sin [ s_0\ln(R/R_t)]$, and the eigenenergies $E_\mathrm{3b}^{(j)}$ are obtained from solving 
\begin{equation}\label{eq:energy}
\mathrm{arg}\ \Gamma\left[\frac{1+is_0-E_\mathrm{3b}^{(j)}/\hbar\omega_0}{2}\right]+s_0 \ln\frac{R_t}{a_\mathrm{ho}}=\mathrm{arg}\ \Gamma\left[1+is_0\right],
\end{equation}
which is evaluated $\mathrm{mod}$ $\pi$.  In the free-space limit of this model ($\omega_0\rightarrow 0$), there exist an infinite number of bound Efimov trimers whose energies and sizes are characterized by the log-periodic geometric scaling \cite{efimov1979low,BRAATEN2006259,BRAATEN2007120,WANG20131,d2017few}:
\begin{equation}\label{eq:efimov}
E_\mathrm{3b}^{(j)}=\frac{E^{(0)}_\mathrm{3b}}{(e^{\pi/s_0})^{2j}}\ \ \text{and}\ \ R_\mathrm{3b}^{(j)}=\sqrt{\frac{2(1+s_0^2)}{3}}\frac{(e^{\pi/s_0})^j}{\kappa_*},
\end{equation}
where $j=0,1,...,\infty$ \cite{collapse}.
In this limit, we choose $R_t$ such that there is a trimer with energy $E^{(0)}_\mathrm{3b}=\hbar^2\kappa_*^2/m\approx 0.051\hbar^2/m r_\mathrm{vdW}^2$, where $\kappa_*$ is the universal three-body parameter found in Ref.~\cite{PhysRevLett.108.263001}.
 
{\it Post-quench dynamics of $\mathcal{C}_3$.}  Given the initial condition in Eq.~(\ref{eq:icpsi}), the solution after quenching is $\Psi(R,{\bf \Omega},t)=\sum_{s,j}c_{s,j}\Psi_{s,j}(R,{\bf \Omega})e^{-i E_\mathrm{3b}^{(j)}t/\hbar}$, with overlaps $c_{s,j}=\langle \Psi_{s,j}|\Psi_0\rangle$ (see Ref.~\cite{SM}.)  This sum runs over all channels, however, the Efimov channel makes the sole contribution to the short-range behavior of three-body correlations at unitarity.  The dominant contributions come from only a few trimers $(E_\mathrm{3b}\leq0)$ and trapped states $(E_\mathrm{3b}>0)$ with eigenenergies comparable in magnitude with $E_n$ \cite{PhysRevA.89.021601}.  At short-range, the relevant behavior of each eigenstate in the Efimov channel is captured by the extensive three-body contact $C_3^{(j)}$, which we have calculated analytically (see Ref.~\cite{SM}).  Intuitively, the dynamical three-body contact density can be written as a superposition of $C_3^{(j)}$ by combining Eqs.~(\ref{eq:g3c3}) and (\ref{eq:psig3link}) and integrating over the hyperangles
\begin{equation}\label{eq:c3quench}
\mathcal{C}_3=\frac{n}{3}\left|\sum_j c_{s_0,j}\times e^{i\phi_j}\sqrt{|C_3^{(j)}|}\ e^{-iE_\mathrm{3b}^{(j)}t/\hbar}\ e^{-\Gamma_j t/2\hbar}\right|^2,
\end{equation}
where $\phi_j\underset{{R}\rightarrow 0}{=}\arg[\Psi_{s_0,j}/\Psi_\mathrm{sc}]$.  Here, we account for three-body losses by utilizing a relation from Refs.~\cite{PhysRevA.86.053633,PhysRevLett.112.110402} to estimate finite widths $\Gamma_j=C_3^{(j)} 4 \hbar\eta/m s_0$ valid in the limit where the inelasticity parameter satisfies $\eta\ll1$.  We assume that this relation is satisfied in the remainder of this Letter.  As a result of the finite width, the time evolution of three-body eigenstates at unitarity is updated time evolution of three-body eigenstates at unitarity is updated $E^{(j)}_\mathrm{3b} \to E_\mathrm{3b}^{(j)}-i\Gamma_j/2$, which leads to a decay of the norm and the form of Eq.~(\ref{eq:c3quench}). 

 \begin{figure}[t!]
\includegraphics[width=8.6cm]{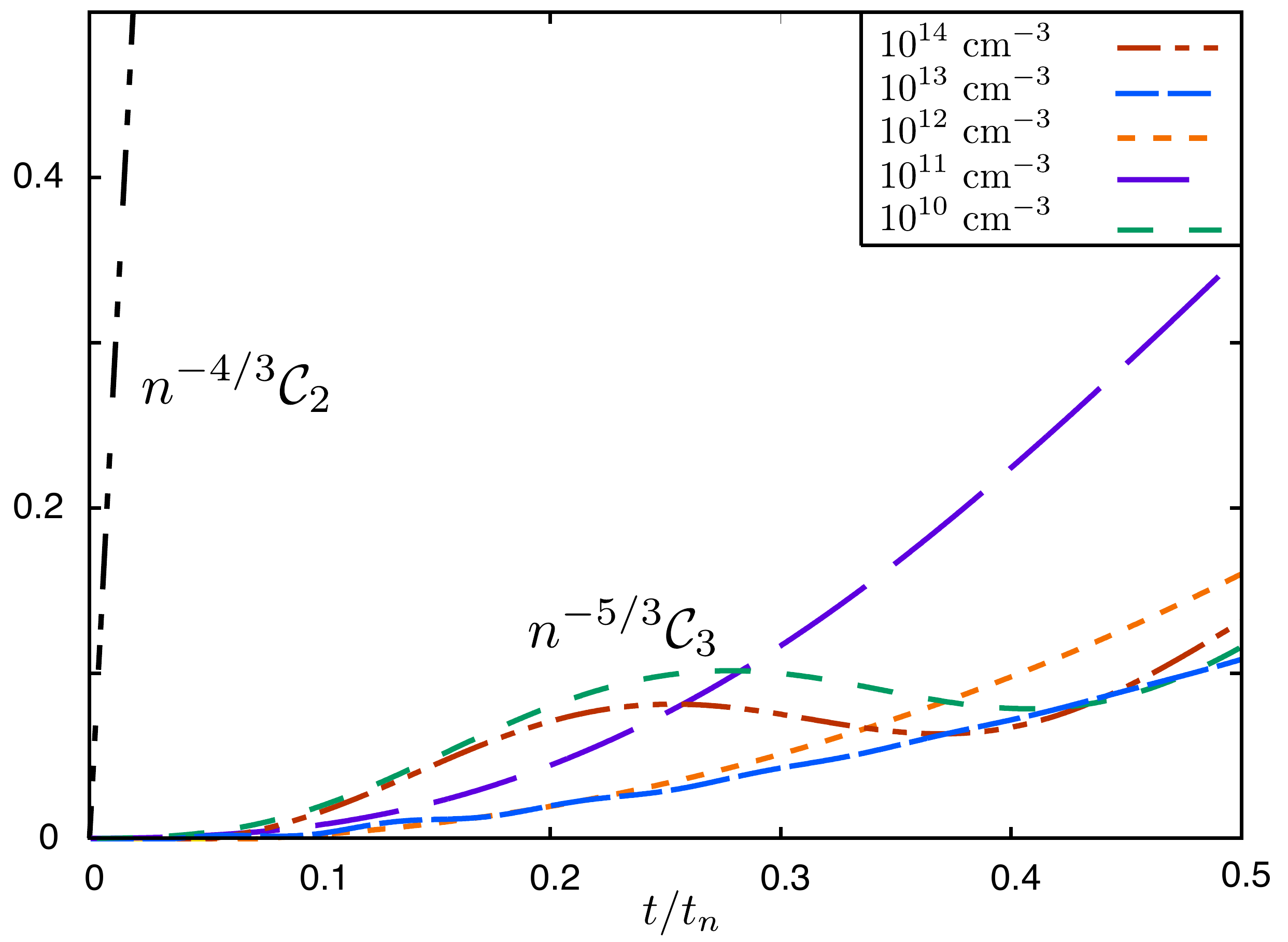}
\caption{(color online) Post-quench dynamics of the two- and three-body contacts over a range of densities.  Evolution of the dimensionless, scaled two-body contact is given by the universal growth rate $n^{-4/3}\mathcal{C}_2=128\pi/(6\pi^2)^{2/3}t/t_n$ from Ref.~\cite{PhysRevA.91.013616}, which quickly increases beyond the plotted range.  This behavior is known and therefore not shown. \label{fig:c2_vs_c3}}
\end{figure}

Hidden in Eq.~(\ref{eq:c3quench}) is a dependence of $\mathcal{C}_3$ on long-range details of the three-body model.  However, the post-quench three-body contact dynamics should depend only on the behavior of the three-body wave function at short distances.  We therefore require our results to be robust to variations of both the trapping parameters  [see Eq.~(\ref{eq:hypeq})] and the arbitrary functional form of $\Psi_0$  [Eq.~(\ref{eq:icpsi})] provided the initial boundary conditions are satisfied.  These criteria are satisfied only at the earliest stages of evolution $t/t_n\lesssim 0.5$.  At later times our model loses physical significance as we expect genuine many-body effects play a role in the correlation dynamics.  These constraints echo the findings of Refs.~\cite{PhysRevA.89.021601,PhysRevA.91.013616}.

The early-time evolution of the two- and three-body contacts in the unitary regime is shown in Fig.~\ref{fig:c2_vs_c3} over a range of densities where contacts have been rescaled by powers of the density into dimensionless form.  Our results for the three-body contact dynamics are specific to $^{85}$Rb, depending on $r_\mathrm{vdW}$, $m$, and $\eta$ for this particular species.  We take $\eta= 0.06$ from experimental measurements in Ref.~\cite{PhysRevLett.108.145305}.  Qualitatively, the contact dynamics agree with the experimental observation in Ref.~\cite{Fletcher377} that the three-body contact develops gradually compared to the two-body contact.  Interpreting $\mathcal{C}_2$ as the number of pairs per (volume)$^{4/3}$ and $\mathcal{C}_3$ as the number of triples per (volume)$^{5/3}$ \cite{TAN20082952,TAN20082971,TAN20082987,PhysRevLett.112.110402}, we find support for the sequential buildup of clusters \cite{kirancomm,KIRA2015185}.  Unlike the early-time behavior of $n^{-4/3}\mathcal{C}_2$ obtained in Refs.~\cite{PhysRevA.89.021601,PhysRevA.91.013616}, the behavior of $n^{-5/3}\mathcal{C}_3$ in Fig.~\ref{fig:c2_vs_c3} varies for different densities.  This is a strong indication of scaling violations in the dynamics of three-body correlations at short-distances (see detailed discussion below.)

\begin{figure}[t]
\includegraphics[width=8.6cm]{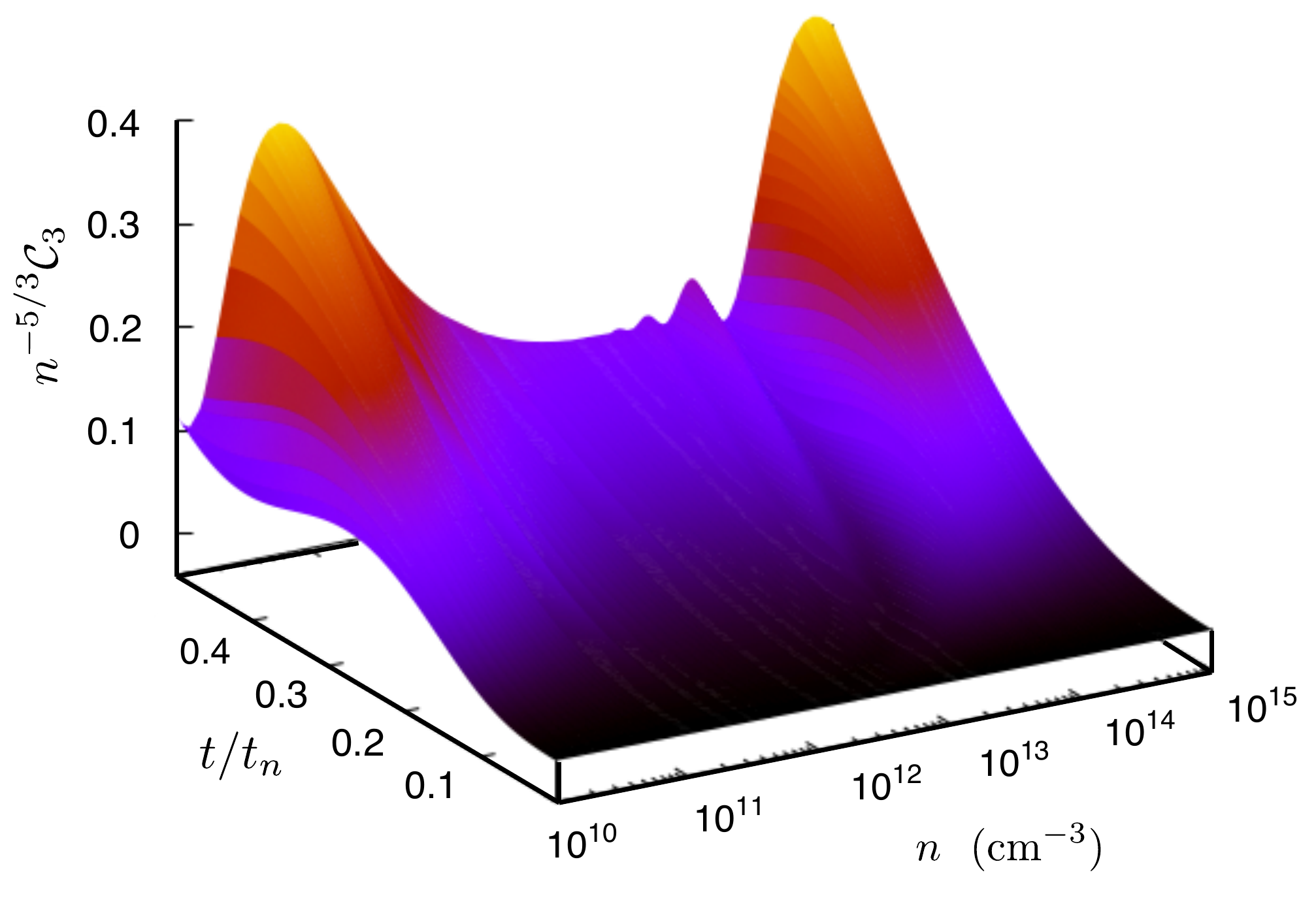}
\caption{\label{fig:surface} (color online) Dynamical surface of $n^{-5/3}\mathcal{C}_3$ over a range of densities.  A ``rippling" effect due to the coherent trimer oscillations occurs as the ``peaks" are approached from lower densities.  This behavior is repeated for densities rescaled by powers of $e^{3\pi/s_0}$.}
\end{figure}

Curiously, for densities $n=10^{10}$ cm$^{-3}$ and $10^{14}$ cm$^{-3}$ the corresponding $n^{-5/3}\mathcal{C}_3$ curves in Fig.~\ref{fig:c2_vs_c3} exhibit a visible oscillation on a timescale shorter than $t_n$.  By eliminating contributions of specific eigenstates to Eq.~(\ref{eq:c3quench}), their origin can be isolated to the Efimov trimer with binding energy nearest $E_n$ satisfying $|E_\mathrm{3b}^{(j)}|\gg E_n$.  Specifically, the oscillation is due to coherences between this trimer and states with energy comparable to $E_n$, resulting in a {\it beating} phenomenon \cite{beats}.  As the energy of this trimer approaches $E_n$ for increasing density, the frequency of the visible oscillations, as well as their amplitude, increases as shown in Fig.~\ref{fig:surface}.  Empirically, we observe that the frequency and damping rate of the oscillations correspond roughly to the frequency $\omega_\mathrm{3b}^{(j)}=E_\mathrm{3b}^{(j)}/\hbar$ and width $\Gamma_j$ of this trimer, respectively.  The trimer oscillations are therefore under-damped and theoretically observable provided $|E^{(j)}_\mathrm{3b}|>\Gamma_j$ obtained whenever  $\eta<s_0/4$ [see Ref.~\cite{SM}.]  Oscillation maxima occur at fixed values of the phase $|E_\mathrm{3b}^{(j)}t|/\hbar= 1.33(11)\pi\ \text{mod}\ 2\pi$.  For the highest and lowest densities in Fig.~\ref{fig:c2_vs_c3}, oscillation is due to the $j=0$ and $j=1$ Efimov trimers, respectively.  Populations of the $j=1$ trimer in the unitary Bose gas were recently observed through a double exponential decay of the molecular gas in Ref.~\cite{klauss2017observation}.  Here, we find additional theoretical evidence for three-body bound-state signatures as coherent beats in the early-time correlation dynamics.  

 {\it Scaling Violations.}  How does Efimov physics alter the density dependence of the early-time evolution of the three-body contact?  The dynamical surface in Fig.~\ref{fig:surface} displays a ``rippling" effect due to the density-independence of the trimer oscillation phase as discussed previously.  There are also a pair of pronounced ``peaks" in Fig.~\ref{fig:surface} due to the variation of the trimer oscillation amplitude with density.  In fact, for fixed values of $t/t_n$ we find identical results for  $n^{-5/3}\mathcal{C}_3$ for densities rescaled by powers of $(e^{\pi/s_0})^{3}$ when plotted as a function of $t/t_n$.  Therefore, the surface in Fig.~\ref{fig:surface} represents only a single log-period, demonstrating that $n^{-5/3}\mathcal{C}_3$ has a {\it discrete} scale-invariance as a direct consequence of Efimov physics.

 \begin{figure}[t]
\includegraphics[width=8.6cm]{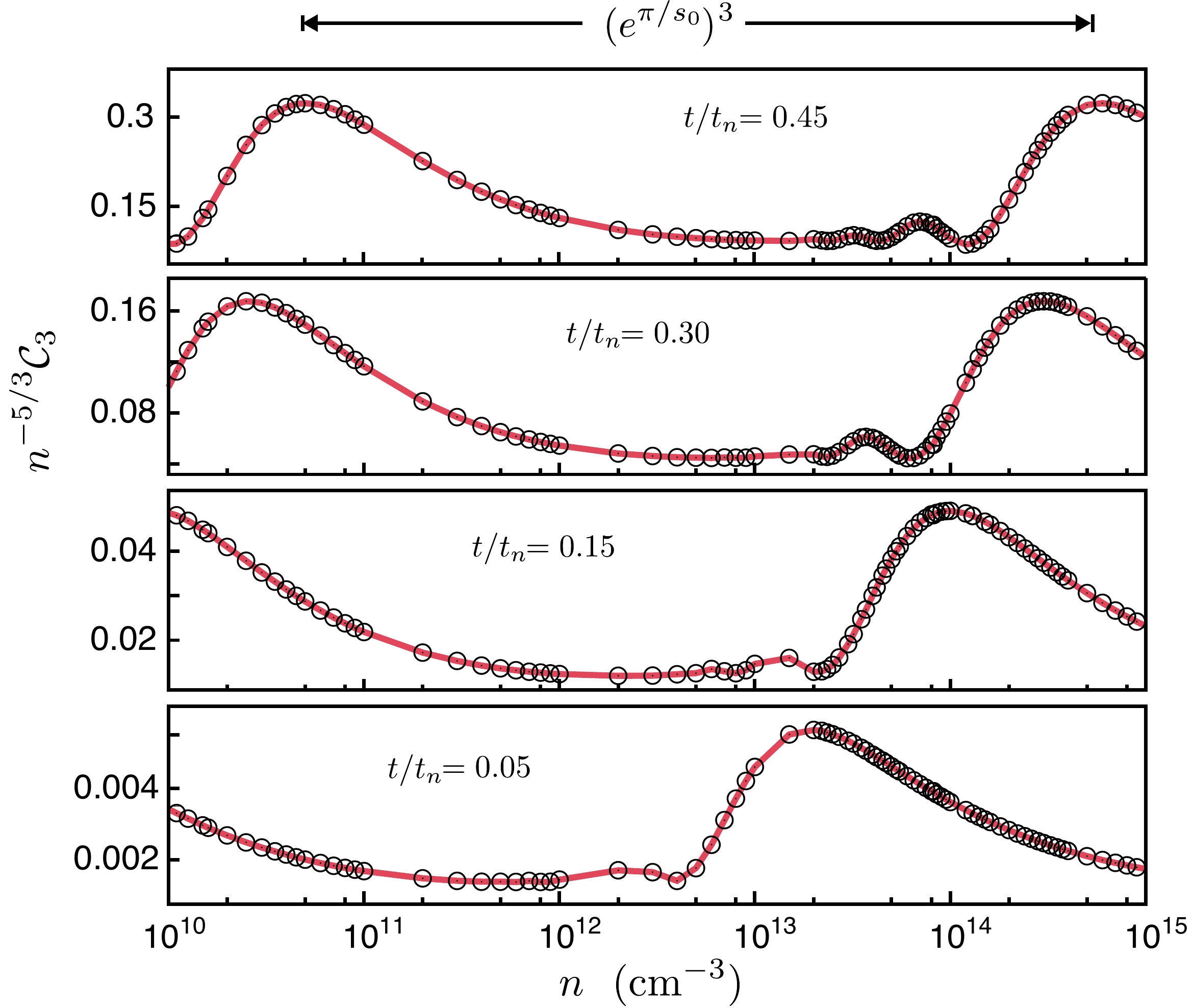}
\caption{\label{fig:envelope} (color online)  Profiles of $n^{-5/3}\mathcal{C}_3$ at fixed $t/t_n$.  Solid lines are guides for the eye connecting data from Fig.~\ref{fig:surface}.}
\end{figure}

With this in mind, we study the envelope of the log-periodic modulation of $n^{-5/3}\mathcal{C}_3$ shown in Fig.~\ref{fig:envelope}.  To characterize the correlation trends, we propose a functional form for the growth of three-body correlations which is quadratic in time to leading order
\begin{equation}\label{eq:fit}
n^{-5/3}\mathcal{C}_3=A\left[1+B\times H(n,\kappa_*,t) \right](t/t_n)^2,
\end{equation}
where $H(n,\kappa_*,t)=H(n e^{3j\pi/s_0},\kappa_*,t)\in[0,1]$ is an unknown log-periodic function reflecting the influence of Efimov physics.  The first term above, proportional to $A$, captures the continuous scale invariant part of the three-body contact, and corresponds to the floor $[\min( n^{-5/3}\mathcal{C}_3)]$ of the curves in Fig.~\ref{fig:envelope}.  In the second term above, the quantity $B $ is the fractional amplitude of the log-periodic modulation $[\max( n^{-5/3}\mathcal{C}_3)-\min( n^{-5/3}\mathcal{C}_3)]/\min( n^{-5/3}\mathcal{C}_3)$ at fixed $t/t_n$ quantifying the violation of the continuous scale invariance.  From fitting our data at early-times $t/t_n\ll1$, we find $A\approx 0.55$, and $B\approx 3.09$.  Therefore the early-time evolution of three-body correlations is {\it in general} poorly-captured by fitting to a universal function with continuous scaling invariance.  In Ref.~\cite{PhysRevLett.112.110402} the restrictive assumption was made that the saturated value of $\mathcal{C}_3$ scales continuously as $n^{5/3}$ with numerically suppressed log-periodic effects.  Over only the limited density range $1.6- 5.5\times 10^{12}$ cm$^{-3}$ fit in Ref.~\cite{PhysRevLett.112.110402}, we find that $H(n,\kappa_*,t)$ is slowly-varying and minimized and hence fitting to a universal scaling law is sufficient only in the immediate vicinity.

Comparing Eq.~(\ref{eq:fit}) to our data, we infer the behavior of $H(n,\kappa_*,t)$, quantifying the violation of scale invariance at particular densities and times.  Within our model, the maximum of this unknown function occurs for densities satisfying 
\begin{equation}\label{eq:criterion}
R_\mathrm{3b}^{(j)}\times k_n= 0.78(2),
\end{equation}
where $j$ is the trimer index as in Eq.~(\ref{eq:efimov}).  When Eq.~(\ref{eq:criterion}) is satisfied, the size of the trimer responsible for the coherent oscillation is comparable to $k_n^{-1}$ and therefore the interparticle spacing.  This results in a correlation enhancement.  Similarly, recent results in Refs.~\cite{PhysRevLett.119.013401,PhysRevA.96.022707} for the Bose polaron problem demonstrated that when the size of Efimov trimers become comparable to the inter particle scaling,  signatures of Efimov physics become visible in the polaron spectrum.   

{\it Conclusion.}  We have studied the early-time dynamics of the three-body contact density for the quenched unitary Bose gas.  The relative growth of the two- and three-body contacts indicates that triples are generated slower than pairs of atoms immediately after the quench.  Efimov physics arises in the dynamics of the three-body contact through a coherent oscillation that is a bound state signature of trimers and through the violation of continuous scale invariance.  It is of considerable interest to extend this analysis to later times beyond the range of our model and to observables depending functionally on the three-body contact.  These investigations may suggest regimes of interest for experiments, which have covered to date \cite{Fletcher377,eigen2017universal,makotyn2014universal,klauss2017observation} a fraction of the complete log-period studied in this Letter although preliminary observations of the decay rate over a wider range of densities display oscillations \cite{klaussthesis}.  With an increase of signal to noise of the measurements in Ref.~\cite{makotyn2014universal}, it may be possible to observe the scale-violations and coherent trimer oscillations predicted in this Letter or through time-resolved RF spectroscopy, as done in Ref.~\cite{Bardon722}.  

The authors thank John Bohn and Servaas Kokkelmans for critical feedback.  J.P.D. acknowledges support from the National Science Foundation (NSF) Grant PHY-1607204 and from the National Aeronautics and Space Administration (NASA).  V.E.C. and J.P.C. acknowledge support from the NSF under Grant PHY-1734006.  V.E.C. is supported also by Netherlands Organisation for Scientific Research (NWO) under Grant 680-47-623.
\bibliographystyle{apsrev4-1}
\bibliography{references}

\begin{thebibliography}{39}
\bibitem [{col()}]{collapse}%
  \BibitemOpen
  \href@noop {} {}\bibinfo {note} {The zero-range model also contains the
  unphysical divergence $(E_\mathrm{3b}^{(j)}\to -\infty$) referred to as the
  Thomas collapse. The trimer states for $j<0$ do not contribute to our model
  (see Ref.~\cite{SM}).}\BibitemShut {Stop}%
\end{thebibliography}
\end{document}


\title{Supplementary Material: ``Dynamics of three-body correlations in quenched unitary Bose gases''}

\author{V. E. Colussi}
\affiliation{Eindhoven University of Technology, PO Box 513, 5600 MB Eindhoven, The Netherlands}
\affiliation{JILA, NIST and Department of Physics, University of Colorado, Boulder, Colorado 80309-0440, USA}
\author{J. P. Corson}
\affiliation{JILA, NIST and Department of Physics, University of Colorado, Boulder, Colorado 80309-0440, USA}
\author{J. P. D'Incao}
\affiliation{JILA, NIST and Department of Physics, University of Colorado, Boulder, Colorado 80309-0440, USA}

\maketitle

\setcounter{equation}{0}
\setcounter{figure}{0}
\setcounter{table}{0}

\renewcommand{\theequation}{S\arabic{equation}}
\renewcommand{\thefigure}{S\arabic{figure}}
\renewcommand{\thetable}{S\Roman{table}}

\section{Three Bosons in a Harmonic Trap}\label{SecI}

In this section, we will give explicit expressions for the solutions for three-bosons in a harmonic trap a unitarity relevant to the main text.  These results were first obtained in Refs.~\cite{PhysRevLett.97.150401,wernerthesis} within the zero-range model.  Here, the interaction potential can be summarized by the Bethe-Peierls contact boundary condition taken at unitary 
\begin{equation}
\Psi({\bf r},{\boldsymbol \rho})\underset{{|\bf r|}\rightarrow 0}{=}\frac{1}{|{\bf r}|}A({\boldsymbol \rho}),
\end{equation}
The unknown function $A$ is determined by solving the three-body Schr\"odinger equation with the above boundary condition.  This problem can be solved analytically for an isotropic harmonic trap using the decomposition $\Psi_{s,j}(R,{\bf \Omega})=\mathcal{N}F^{(s)}_j(R)\phi_s({\Omega})/R^2\sin2\alpha,$ given in the main text.  For s-wave interactions, the hyperangular eigenstates solve the set of equations 
\begin{eqnarray}
&&-\phi''(\alpha)=s^2\phi(\alpha),\\
&&\phi(\pi/2)=0,\\
&&\phi'(0)+\frac{8}{\sqrt{3}}\phi(\pi/3)=0.
\end{eqnarray}
These equations define a transcendental equation for $s$
\begin{equation}
\frac{8}{\sqrt{3}}\sin\left(\frac{s\pi}{6}\right)-s\cos\left(\frac{s\pi}{2}\right)=0.
\end{equation}
This transcendental equation has an infinite number of real solutions and a single imaginary solution $s=is_0$ where $s_0\approx 1.00624$ labelling the Efimov channel.  Hyperangular eigenstates can be factorized $\phi_{s}({\bf \Omega})=\left(1+\hat{P}_{13}+\hat{P}_{23}\right)\sin\left[s\left(\frac{\pi}{2}-\alpha\right)\right]/\sqrt{4\pi}$,in terms of operators $\hat{P}_{ij}$ that permute particles $i$ and $j$.  In order for the problem to be Hermitian \cite{danilov,morse,PhysRevLett.97.150401}, there is an additional boundary condition that the hyperradial eigenfunctions must satisfy in the Efimov channel, $\lim_{R\to 0}F^{(s_0)}_j(R)\propto\sin [ s_0\ln(R/R_t)].$  The hyperradial eigenstates are \cite{PhysRevLett.97.150401}  
\begin{equation}
F_j^{(s_0)}(R)=\frac{a_\mathrm{ho}}{R}W_{E_\mathrm{3b}^{(j)}/2\hbar\omega_0,is_0/2}(R^2/a_\mathrm{ho}^2),
\end{equation}
where $W_{n,m}$ is the Whittaker function.  As $R\to 0$, the limiting form of the hyperradial eigenstates to leading order is \cite{olver2010nist}
\begin{equation}
F_j^{(s_0)}(R)=2i R_t^{-is_0}\frac{\Gamma(is_0)}{\Gamma\left(\frac{1+is_0-E_\mathrm{3b}^{(j)}/\hbar\omega_0}{2}\right)}\sin\left[s_0\ln\frac{R}{R_t}\right],
\end{equation}
which oscillates log-periodically as required by the additional boundary condition in the Efimov channel.  

The hyperradial solutions are trapped and therefore normalizable.  The normalization integrals can be found in Ref.~\cite{gradshteyn2014table,wernerthesis}, which we reproduce here
\begin{eqnarray}
\langle F_j^{s_0}|F_j^{s_0}\rangle&=&\frac{1}{a_\mathrm{ho}^2}\int dR\ R\ |F_j^{(s_0)}(R)|^2,\nonumber\\
 &=&\frac{\pi\text{Im}\psi\left(\frac{1+E_\mathrm{3b}^{(j)}/\hbar\omega_0+is_0}{2}\right)}{\sinh(s_0\pi)\left|\Gamma\left(\frac{1+E_\mathrm{3b}^{(j)}/\hbar\omega_0+is_0}{2}\right)\right|^2},\nonumber\\
\end{eqnarray}
where $\psi$ is the digamma function.  Normalization of the hyperrangular eigenstates was first given in Ref.~\cite{PhysRevA.83.063614} which we reproduce for completeness
\begin{eqnarray}
\langle \phi_s|\phi_s\rangle&=&\int d{\bf \Omega} |\phi_s(\Omega)|^2,\nonumber\\
&=&\frac{3\pi}{2s}\sin\left(\frac{s\pi}{2}\right)\left[\cos\left(\frac{s\pi}{2}\right)\right.\nonumber\\
&&-\left.\frac{s\pi}{2}\sin\left(\frac{s\pi}{2}\right)-\frac{4}{3\sqrt{3}}\cos\left(\frac{s\pi}{6}\right)\right],\nonumber\\
\end{eqnarray}
where 
\begin{equation}
\int d{\bf \Omega} =\frac{1}{4}\int_0^{\pi/2}d\alpha\ \int d{\bf\hat{\rho}}\int d {\bf\hat{r}}.
\end{equation}
The overall normalization factor $\mathcal{N}$ for each eigenstate is
\begin{equation}
\mathcal{N}_{s_0,j}^2=\frac{1}{a_\mathrm{ho}}\frac{1}{\ 3^{3/2}\langle F_j^{(s_0)}|F_j^{(s_0)}\rangle\ \langle\phi_{s_0}|\phi_{s_0}\rangle},
\end{equation}
where the factor of $3^{3/2}$ arises due to the Jacobian $D({\bf r_1},{\bf r_2},{\bf r_3})/D({\bf c}_{123},R,{\bf \Omega})=3\sqrt{3}$.
\section{Normalization of the Initial Wave Function}
In the calibration scheme presented in the main text, the initial wave function must satisfy a pair of boundary conditions along with the usual normalization constraint $\langle \Psi_0|\Psi_0\rangle=1$.  This constrains the normalization factor $A$ for the initial three-body wave function $\Psi_0(R,{\bf \Omega})$, which must satisfy
\begin{equation}
A=\frac{1}{(3\pi^2)^{3/4}\sqrt{B_1^6+12B_1^{10}/B_2^4-6B_1^8/B_2^2}}.
\end{equation} 
\section{Overlap of Initial wave function and Eigenstates at unitarity}
The overlaps between the eigenstates at unitarity and the initial three-body wave function are calculated from the projection $c_{s_0,j}=\langle \Psi_{s_0,j}|\Psi_0\rangle$ as
\begin{equation}\label{eq:overlap}
c_{s_0,j}=3^{3/2} \int dR\ R^5 \int d\Omega\ \Psi_0(R,{\bf \Omega})\Psi_{s_0,j}^*(R,{\bf \Omega}).
\end{equation}
The hyperrangular integration can be easily performed and the total expression for the overlap reduces to
\begin{equation}
|c_{s_0,j}|^2=P(s_0)\frac{|\langle F_j^{(s_0)}|F_0\rangle|^2}{\langle F_j^{(s_0)}|F_j^{(s_0)}\rangle \langle F_0|F_0\rangle}, 
\end{equation}
where $|F_0\rangle$ is the hyperradial component of $|\Psi_0\rangle$. For general $s$, the total contribution to the norm of each channel is given by $\sum_{n}|c_{s,n}|^2=P(s)$ which was first obtained in Ref.~\cite{wernerthesis} as
\begin{eqnarray}\label{eq:overlapsdetail}
P(s)&=&\left[\frac{s\pi}{2}\sin\left(\frac{s\pi}{2}\right)-\cos\left(\frac{s\pi}{2}\right)+\frac{4}{3\sqrt{3}}\cos\left(\frac{s\pi}{6}\right)\right]^{-1}\nonumber\\
&&\times\frac{96 s\sin\left(\frac{s\pi}{2}\right)}{\pi(s^2-4)^2}.
\end{eqnarray}
We calculate the lengthy analytic expression for $\langle F_j^{(s_0)}|F_0\rangle$ in Eq.~\ref{eq:overlapsdetail} which can be found in Ref.~\cite{gradshteyn2014table} and is required to reproduce the results in the main text.
\begin{equation}
\langle F_j^{(s_0)}|F_0\rangle=\frac{1}{a_\mathrm{ho}^3}\left(\frac{1}{2}\ \Xi\left[\frac{1}{2},j\right]-\frac{1}{2B_2^2}\ \Xi\left[\frac{3}{2},j\right]\right),
\end{equation}
where
\begin{eqnarray}
\Xi[\alpha,j]&=&\frac{\left|\Gamma\left(\alpha+is_0/2+3/2\right)\right|^2}{\Gamma\left(\alpha+2-E_\mathrm{3b}^{(j)}/2\hbar\omega_0\right)}\times \left(\frac{1}{a_\mathrm{ho}^2}\right)^{(is_0+1)/2}\nonumber\\
&\times& \text{}_2F_1\left(a,\ b;\ c,\ z\right)\nonumber\\
&\times& \left(\frac{1}{2B_1^2}+\frac{1}{2 a_\mathrm{ho}^2}\right)^{-is_0/2-3/2-\alpha}.
\end{eqnarray}
The function $\text{}_2F_1$ is the Gauss hypergeometric function \cite{abramowitz1964handbook} with arguments
\begin{eqnarray}
a&=&3/2+\alpha+is_0/2,\\
b&=&\left(is_0-E_\mathrm{3b}^{(j)}/\hbar\omega_0+1\right)/2,\\
c&=&\alpha+2-E_\mathrm{3b}^{(j)}/2\hbar\omega_0,\\
z&=&\frac{1/B_1^2-1/a_\mathrm{ho}^2}{1/B_1^2+1/a_\mathrm{ho}^2}.
\end{eqnarray}
\begin{figure}[t!]
\includegraphics[width=8.6cm]{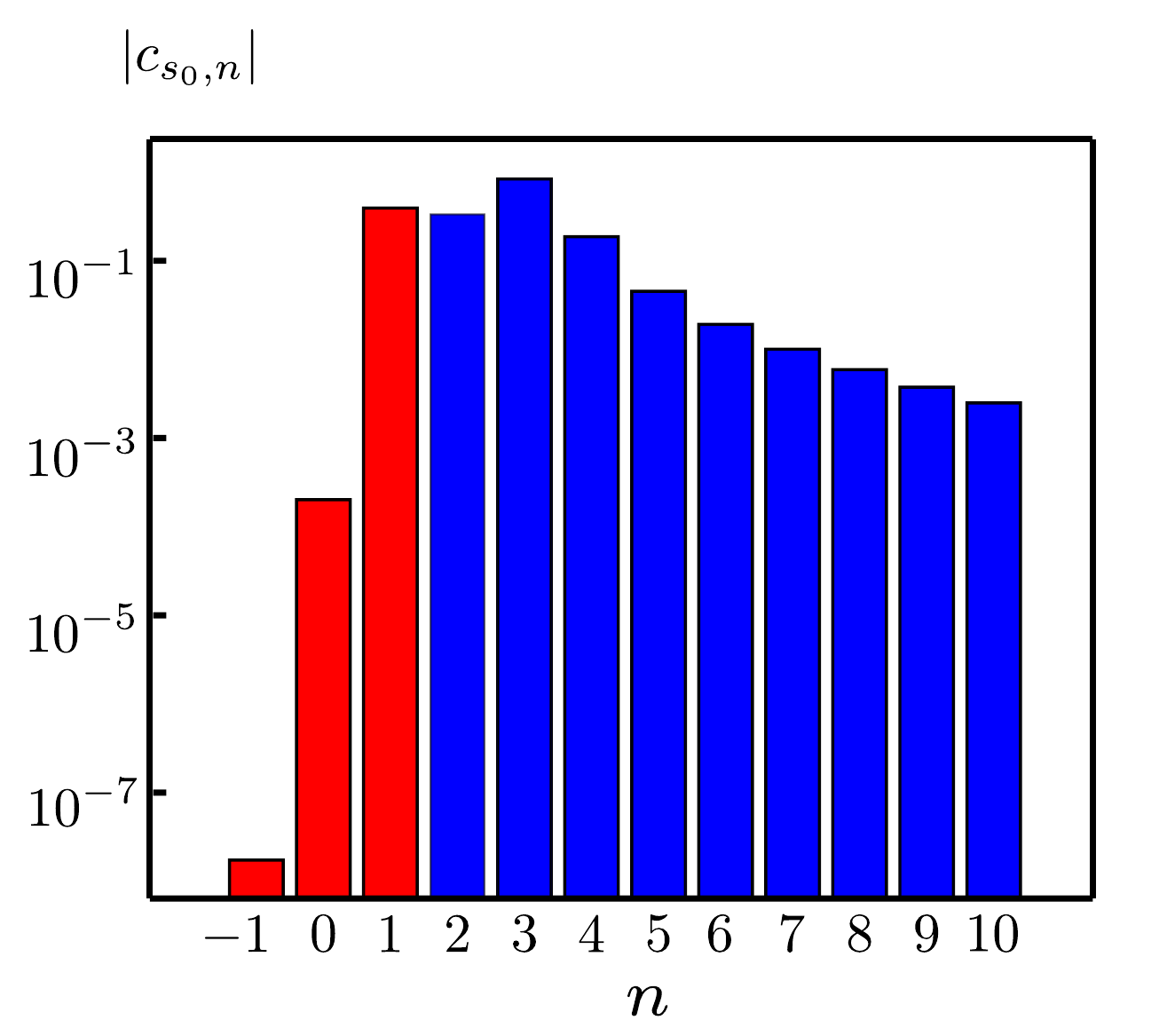}
\caption{\label{fig:overlaps} Modulus of the overlaps for eigenstates in the Efimov channel for $n=1\times 10^{12}$ $\text{cm}^{-3}$ with $a_\mathrm{ho}=B_1\approx 0.6009n^{-1/3}$.  Overlaps corresponding to negative energy eigenstates are marked in red, and the positive energy trapped states are marked in blue.}
\end{figure}
The limiting cases $|E_\mathrm{3b}^{(j)}|\to\infty$ of $\langle F_j^{(s_0)}|F_0\rangle$ require asymptotic expressions for the hypergeometric function $ \text{}_2F_1\left(a,\ b;\ c,\ z\right)$ when $b$, $c\to\pm\infty$.  Such expressions can be found tabulated in Ref.~\cite{doi:10.1142/S0219530514500389}, however we do not reproduce the lengthy result here.  Rather, numerical results for the overlaps are shown in Fig.~\ref{fig:overlaps}, where convergence in the limiting cases is evident.  
\section{Extensive three-body contact for three bosons in an isotropic trap at unitarity}
The dynamics of the three-body contact density $\mathcal{C}_3$ can be written as a superposition of the extensive three-body contacts $C_3^{(j)}$ for each of the eigenstates in the Efimov channel at unitarity.  We have developed a simple method of extracting the $C_3^{(j)}$ from the short-range behavior of the three-body wave function.  

Here we outline this method for calculating the extensive three-body contact for each trapped three-body eigenstate in the Efimov channel at unitarity.  For three bosons, we begin with the general relation between the non-normalized three-body correlation function $G^{(3)}$ and the three-body wave function
\begin{equation}
G^{(3)}(R,{\bf \Omega})=3!|\Psi(R,{\bf \Omega})|^2,
\end{equation}
where the factor $3!$ arises from the total number of permutations for three indistinguishable particles.  A general expression derived in Ref.~\cite{PhysRevA.86.053633} relates the short-range behavior of $G^{(3)}$ with the three-body contact
\begin{equation}
G^{(3)}(R,{\bf \Omega})=|\Psi_\mathrm{sc}(R,{\bf \Omega})|^2\frac{8}{\sqrt{3}s_0^2}C_3
\end{equation}
Therefore, the relative three-body wave function for a trapped eigenstate is related with the extensive three-body contact
\begin{equation}
 \frac{|\Psi_{s_0,j}(R,{\bf \Omega})|^2}{\langle F_j^{(s_0)}|F_j^{(s_0)}\rangle\langle \phi_{s_0}|\phi_{s_0}\rangle}=\left|\Psi_\mathrm{sc}(R,{\bf \Omega})\right|^2\frac{4}{s_0^2}C_3^{(j)}.
 \end{equation}
 After integrating over the hyperangles, $C_3^{(j)}$ can be obtained as a simple proportionality constant in the limit $R\to 0$ as
 \begin{equation}\label{eq:c3}
 C_3^{(j)}=\frac{s_0^2}{a_\mathrm{ho}^2\langle F_j^{(s_0)}|F_j^{(s_0)}\rangle}\left|\frac{\Gamma(is_0)}{\Gamma\left(\frac{1+is_0-E_\mathrm{3b}^{(j)}/\hbar\omega_0}{2}\right)}\right|^2.
\end{equation}
Numerically, in the free space limit ($\omega_0\to 0$), we find good agreement between $C_3^{(j)}$ and the free space trimer energies.  This is expected from the operational definition 
\begin{equation}
\left(\kappa_*\frac{\partial E_\mathrm{3b}^{(j)}}{\partial \kappa_*}\right)_a=-\frac{2\hbar^2}{m} C_3^{(j)},
\end{equation}
given in Ref.~\cite{PhysRevLett.106.153005}.  

 %
 %
 \bibliographystyle{apsrev4-1}
\bibliography{references_SM}